\documentclass[prb,twocolumn,floatfix,showpacs]{revtex4}
\usepackage{graphicx,subfigure,amsmath,colordvi,times}
 \usepackage[usenames]{color}
 \usepackage{multirow,bm}
 
\begin{document}
 \title{Excitonic Superfluidity and Screening in Electron-Hole Bilayer Systems }
\author{D. Neilson$^{1}$, A. Perali$^1$,  A.R. Hamilton$^2$}
\affiliation{$^1$Universit\`a di Camerino, 62032 Camerino (MC), Italy\\
$^2$School of Physics, University of New South Wales, Sydney 2052, Australia}

\begin{abstract}
Superfluidity in electron-hole bilayers in graphene and GaAs has been predicted theoretically many times but not yet definitively observed.  A key controversy is the correct approximation for the screening of the Coulomb interaction for the pairing.  Mean-field theories using different approximations for the screening lead to diametrically contradictory predictions for superfluidity.  We test these different approximations against Diffusion Quantum Monte-Carlo results and find good agreement with the mean-field theory that uses screening in the superfluid state, but large discrepancies with other approximations for screening.  This mean-field theory predicts no superfluidity in existing devices, provides pointers for new devices to generate the superfluidity, {and, very importantly, it permits} calculations for complicated lattices at finite temperatures, impractical in Monte-Carlo.  
%783 characters
\end{abstract}
\pacs{71.35.-y, 73.21.-b, 73.22.Gk, 74.78.Fk}
\maketitle

There have been intense efforts to observe excitonic superfluidity in bilayer electron-hole systems.  These include double quantum wells in GaAs-AlGaAs heterostructures, double graphene monolayers, and hybrid graphene-GaAs structures.\cite{Sivan,Croxall,Lilly,Gorbachev,Polini2012}   With a Coulomb  interaction to generate pairing between electrons and holes, there are  predictions of room-temperature superfluidity in such systems.\cite{ZhangMin2008}   But despite 
ultra-high quality materials, and insulating barriers as thin as $1$ nm, the superfluid has not yet been definitively observed,
except in the quantum Hall regime with an external magnetic field, where the physics is quite different.\cite{EisensteinMacDonald2004}

{An important theoretical controversy} involves the nature and effectiveness of the screening of the long range Coulomb electron-hole interaction driving the superfluid pairing.    
There have been suggestions that  extremely strong screening will completely suppress superfluidity in graphene double layers,\cite{KE2008}  but other calculations that treat screening differently 
arrive at a diametrically opposite conclusion that high temperature superfluidity in this system should indeed be possible.\cite{ZhangMin2008,BMSMcomment,Lozovik2012,Sodermann2012,Perali2013}
There is {disagreement} in the literature as to whether (a) the pairing interaction should be unscreened (US),\cite{Pieri,ZhangMin2008}  or (b) to work with a screened pairing interaction appropriate for a normal state (NS),\cite{GortelSwier96,KE2008,DasSarma} or (c)
to start with a superfluid state generated by the unscreened pairing interaction and then self-consistently screen the pairing interaction by  carriers in the superfluid coherent state (SS).\cite{BMSMcomment,Lozovik2012,Sodermann2012,Perali2013} 
These different approaches predict {dramatically different properties} for the electron-hole system.

Recently, an upturn in the Coulomb drag has been reported as $T$ goes to zero in GaAs and in graphene-GaAs hybrid heterostructures.\cite{Croxall,Lilly,Pelligrini2014}  This may be a precursor of electron-hole superfluidity.  The upturn occurs at low temperatures that are in agreement with only some of the theoretical predictions.

In this work we are able to resolve the controversy on the correct mean-field approximation for screening by comparing the different approximations against ground state results from highly accurate Diffusion Quantum Monte Carlo (DQMC) calculations.   DQMC serves as a benchmark against which approximate theories may be compared.\cite{Foulkes}  The accuracy of DQMC results has been confirmed in related systems by agreement within a few percent with experimental measurements of the BCS-BEC crossover of ultra-cold strongly interacting fermions.\cite{Carlsonetal} 

The exciton superfluid condensate fraction will serve here as the calibration measure for the approximate mean-field theories of screening.   The condensate fraction is a fundamental ground state property, extensively used experimentally and theoretically to characterize the different regimes of pairing in systems of ultra-cold strongly interacting fermions.\cite{Salasnich2005}  Recently DQMC has been used to study condensation in the electron-hole bilayer \cite{Maezono2013} (see also Ref.\ \onlinecite{DPRS2002}), including calculating the condensate fraction. 
The system investigated in Ref.\ \onlinecite{Maezono2013} is a symmetric single-valley electron-hole bilayer with quadratic energy bands.   We compare the DQMC superfluid condensate fraction properties with mean-field calculations for the same system using the three approaches for screening referred to above, (US), (NS), and (SS).   	

Another open theoretical problem is how to correctly deal with vertex corrections in the mean-field calculations.\cite{Pietronero,DasSarma} 
Since DQMC includes not only full screening but also vertex corrections and two-body correlations, while all the mean-field screening approaches omit vertex corrections and intralayer correlations, it means that comparisons of the predicted condensate fractions can also provide new information on the importance in the excitonic superfluid state    
of the vertex corrections and intralayer correlations.   

We describe the electron-hole bilayer system by the grand-canonical Hamiltonian,
\begin{eqnarray}
&&{\cal{H}}=\sum_{\mathbf{k},\ell} \xi_{\mathbf{k}}
c^{\dagger}_{\mathbf{k} \ell} c_{\mathbf{k} \ell} 
\nonumber\\
&&+\frac{1}{2\Omega}
\sum_{\mathbf{k},\mathbf{k'},\mathbf{q},\ell \neq \ell'} V^0_{|\mathbf{k}-\mathbf{k'}|}
c^{\dagger}_{\mathbf{k}+\frac{\mathbf{q}}{2} \ell}
c^{\dagger}_{-\mathbf{k}+\frac{\mathbf{q}}{2} \ell'}
c_{-\mathbf{k'}+\frac{\mathbf{q}}{2} \ell'}
c_{\mathbf{k'}+\frac{\mathbf{q}}{2} \ell}\ \ \ \ \ \ 
\label{Grand-canonical-Hamiltonian}
\end{eqnarray}
$\mathbf{k}$, $\mathbf{k'}$, and $\mathbf{q}$ are
two-dimensional wave vectors in the layers, $\Omega$ is the
quantization area, 
$c^{\dagger}_{\mathbf{k}
\ell}$ ($c_{\mathbf{k} \ell}$) are the creation (destruction)
operators for electrons ($\mathrm{e}$) and holes ($\mathrm{h}$)
distinguished by $\ell=(\mathrm{e},\mathrm{h})$, and
the quadratic band dispersion for the electrons and holes of equal effective mass $m^{\star}$ are $\xi_{\mathbf{k}}= \mathbf{k}^{2}/(2 m^{\star}) - \mu$, with $\mu$ the equal electron and hole chemical potentials.
Spin quantum numbers are not shown.  $V^0_q=v_q\mathrm{e}^{-qd}$ is the bare Coulomb interaction between electrons and holes separated by a barrier of thickness $d$ and dielectric constant $\kappa$, with   
$v_q=-2\pi e^2/(\kappa q)$.

The effective electron-hole interaction $V_{q}$ in the unscreened case (US), and with RPA screening in the normal state (NS) is
\begin{eqnarray}
\!\!\!\!V^{\mathrm{(US)}}_{q}\!\!\!\!&=&\!\!\!V^0_q
%\nonumber
\\
\!\!\!\!V^{\mathrm{(NS)}}_{q}\!\!\!\!&=&\!\!\!\frac{V^0_q}{1+2v_q\Pi_0^{(NS)}(q)+\left(v_q\Pi_0^{(NS)}(q)\right)^2\!\!\left[1-\mathrm{e}^{-2qd}\right]}
\label{VSS}
\end{eqnarray}
where 
$\Pi_0^{(NS)}(q)$ is the  polarizability in the normal state within a layer.  
In the (SS) approach, calculations start with the coherent state generated by the unscreened interaction $V^0_q$, and the pairing interaction is then screened within the RPA by carriers in the superfluid coherent state which spans  the two layers,
\begin{eqnarray}
\!\!\!\!V^{\mathrm{(SS)}}_{q}\!\!\!\! &=& \!\!\!\frac{V^0_q}{1+2v_q\Pi_0^{(SS)}(q)+\left(v_q\Pi_0^{(SS)}(q)\right)^2\!\!\left[1-\mathrm{e}^{-2qd}\right]}
\end{eqnarray}
where 
  $\Pi_0^{(SS)}(q)=\Pi_0^{(n)}(q)+ \Pi_0^{(a)}(q)$, with  $\Pi_0^{(n)}(q)$ and $\Pi_0^{(a)}(q)$ the normal and anomalous polarizabilities in the superfluid state.\cite{Perali2013}  

The $T=0$ mean field equations for the (s-wave) gap $\Delta_{\mathbf{k}}$ and chemical potential $\mu$ for equal carrier densities $n$ are,
\begin{equation}
\Delta_{\mathbf{k}} = - \frac{1}{\Omega} \sum_{\mathbf{k'}} V_{|\mathbf{k}-\mathbf{k'}|}
\frac{\Delta_{\mathbf{k'}}}{2 E_{\mathbf{k'}}}
;\ \ \ \ \ 
n = \frac{2}{\Omega} \sum_{\mathbf{k}}
 ( 1 - \xi_{\mathbf{k}}/E_{\mathbf{k}})
,
\label{Delta-eqn}
\end{equation}
where $E_{\mathbf{k}} = \sqrt{\xi_{\mathbf{k}}^2 +
\Delta_{\mathbf{k}}^{2}}$.  
%The carrier density $n$ of the layers is equal.
   
%%%%%%%%%%%%%%%%%%%%%%%%%%%%%%%%%%%%%%%%%%%%%%%%%%%%%%%%%%%%%%
 \begin{figure}
\includegraphics[width=0.48\textwidth]{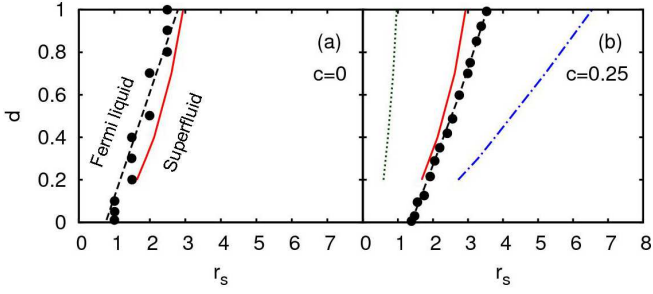}
\caption{(a) Superfluid-normal phase diagram at zero temperature. 
Axes are layer separation $d$ and inter-particle spacing $r_s$.
Condensate fraction phase boundary $c=0$ for DQMC taken from Ref.\ \onlinecite{Maezono2013} (dashed black curve with filled circles), 
and for superfluid state screened  interaction (SS) (solid red line).%\\ 
(b) Condensate fraction $c=0.25$ contour  line for DQMC,
for screened interaction in the superfluid state (SS),
for unscreened (US) (dotted green line),   
and for screened in the normal state (NS) (dash-dot blue line). 
}
 \label{Critical-density}
 \end{figure}
%%%%%%%%%%%%%%%%%%%%%%%%%%%%%%%%%%%%%%%%%%%%%%%%%%

Figure \ref{Critical-density}(a) shows the superfluid-normal phase diagram at zero temperature.  The axes are the barrier thickness $d$, and the density, represented in terms of the average particle spacing $r_s$, both in units of the effective Bohr radius.
% as defined in Ref.\ \onlinecite{Maezono2013}, $a_{0}^{\star}=\kappa \hbar^2/(m^*_r e^2)$, with $m^*_r$ the reduced mass.   
At high densities, the DQMC predicts a negligible exciton condensate fraction, then at a threshold around $r_s \sim 1$--$2$, the condensate fraction abruptly jumps to values of order unity.  The DQMC $c=0$ contour (dashed black curve with filled circles), represents the boundary that separates the superfluid phase from the normal Fermi liquid.  This contour is reproduced from Fig.\ 3, and Fig.\ 1 of the Supplementary Material of Ref.\ \onlinecite{Maezono2013}.     

The (SS) mean-field calculation gives a jump in the condensate fraction similar to the jump predicted by DQMC, and we see in Fig.\ \ref{Critical-density}(a) that this position of the (SS) $c=0$ contour (solid red line) 
reproduces the DQMC normal-superfluid phase boundary very well.   In contrast, the (US) and (NS) mean-field approximations  show no discontinuous jump in the condensate fraction, predicting instead a continuous exponential growth in the condensate fraction with increasing $r_s$, starting at zero in the $r_s=0$ limit.    

Since the (US) and (NS) approaches have no threshold for condensate formation, in Fig.\ \ref{Critical-density}(b) we instead  compare the point in DQMC  and in the three mean-field approximations at which $c$ reaches $c=0.25$.  The DQMC $c=0.25$ contour line is reproduced from  Fig.\ 1 of Ref.\ \onlinecite{Maezono2013} Supplementary Material.
The $c=0.25$ contour line from the (SS) approach 
%with the superfluid state screened interaction 
is again in good agreement with the $c=0.25$ contour line from DQMC.  In contrast, the  $c=0.25$ contour lines from the (US) and  (NS) approaches are seen to lie well to the left and well to the right of the DQMC contour line, respectively.  

%%%%%%%%%%%%%%%%%%%%%%%%%%%%%%%%%%%%%%%%%%%%%%%%%%%%%%%%%%%%%%
 \begin{figure}
\includegraphics[width=0.48\textwidth]{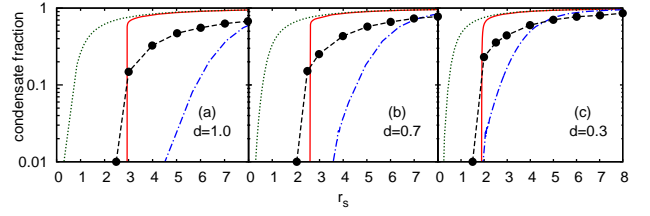}
\caption{Condensate fraction $c$ as function of $r_s$  for barrier thickness $d$ as labeled. %(a) $d=1$, (b) $d=0.7$, (c) $d=0.3$. 
DQMC (Ref.\ \onlinecite{Maezono2013}) (dashed black curve with filled circles); 
unscreened (US) (dotted green line); screened in superfluid state (SS) (solid red line); screened in  normal state (NS) (dash-dot blue line). 
}
 \label{Cond-Frac}
 \end{figure}
%%%%%%%%%%%%%%%%%%%%%%%%%%%%%%%%%%%%%%%%%%%%%%%%%%

We now compare the functional dependence  on $r_s$ of the condensate fractions at fixed $d$ for DQMC and the mean-field approximations.  
Figure \ref{Cond-Frac}(a) compares the respective condensate fractions at $d=1.0$.   The DQMC condensate fraction is reproduced from Fig.\ 2 of Ref.\ \onlinecite{Maezono2013}.  For $r_s<2.5$,  DQMC predicts a negligible exciton condensate fraction.\cite{footnote}  Then at $r_s\simeq 2.5$, the condensate fraction discontinuously jumps from zero to $c\sim 0.2$. Thus for $d=1.0$, the position of the DQMC  normal-superfluid phase boundary is at $r_s\simeq 2.5$.     
The corresponding condensate fraction from the (SS) approach shows a similar discontinuous jump at $r_s\simeq 3$, also from exponentially small values to $c\sim 0.4$.   The (US) and (NS) condensate fractions exhibit no jump, but instead grow smoothly and exponentially from zero in the $r_s=0$ limit.  Thus, as we have noted, the normal-superfluid  phase boundary predicted by DQMC, does not exist for the (US) and (NS) approaches.
% to screening.  

When we take $r_s$ above the onset value, the DQMC and (SS) condensate fractions in Fig.\ \ref{Cond-Frac}(a) are of order unity and increase rapidly.  However, the (SS) condensate fraction grows significantly faster than the DQMC condensate fraction.  This discrepancy, which does not exceed a factor of two,    
is associated with the formation of biexcitons in the DQMC calculation, an effect of $4$-particle correlations which are absent in  mean-field theories.   For large $r_s$, biexciton formation becomes significant at the expense of exciton formation, and this has the effect of significantly reducing the DQMC exciton condensate fraction.\cite{Maezono2013,DPRS2002}    
Figures \ref{Cond-Frac}(b) and (c) show similar overall results for $d=0.7$ and $0.3$.   We see a similar level of agreement between  DQMC and (SS) results maintained over the range $0.3\leq d\leq 1$ .  The DQMC,  (SS), and (US) results all have a weak dependence on $d$, but the (NS)  curve moves to sharply smaller $r_s$ values with decreasing $d$.

{Recalling that DQMC is a benchmark for ground state properties and includes full dynamic screening, the full vertex corrections, and the intra- and interlayer two-body density correlations, we conclude that the comparisons with  DQMC}  in Figs.\ \ref{Critical-density} and \ref{Cond-Frac} strongly indicate that the (SS) approach is the most reliable mean-field approximation for screening in the presence of a superfluid.
	
A central  consideration for experiments is the expected transition temperature $T_c$ for the superfluid, since a large superfluid condensate fraction at $T=0$ is not of practical interest if $T_c$ is so low that it is experimentally inaccessible.   $T_c$ cannot be directly determined from ground state properties because in two dimensions $T_c$ is not linearly related to the value of the $T=0$ gap $\Delta$,\cite{KT1973}  but nevertheless a large value of $\Delta$ through strong pairing is an essential prerequisite for a high $T_c$.  For example, Ref.\  \onlinecite{KE2008} concluded from their determination of an extremely weak pairing energy scale  in double monolayer graphene, that any superfluid transition would occur at impractically low $T_c$.   At present there exists no DQMC calculation of the superfluid gap $\Delta$, but now that the (SS) mean-field approach has been validated against highly accurate DQMC calculations, we can use the (SS) approach to predict $\Delta$.  
  
  %%%%%%%%%%%%%%%%%%%%%%%%%%%%%%%%%%%%%%%%%%%%%%%%%%%%%%%%%%%%%%
 \begin{figure}
\includegraphics[width=0.4\textwidth,height=0.25\textwidth]{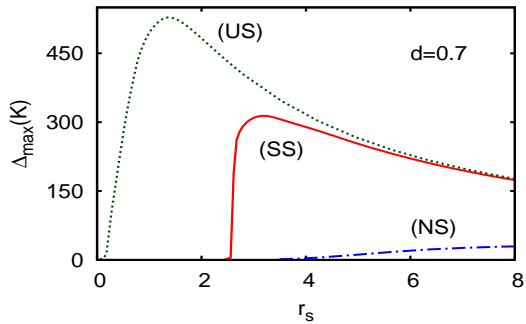}
\caption{Superfluid gap $\Delta_{\mathrm{max}}$ at $T=0$ as a function of $r_s$, calculated for a Coulomb electron-hole pairing interaction which is unscreened (US) $\Delta_{\mathrm{max}}^{\mathrm{(US)}}$ (dotted green line); screened in the superfluid state (SS) $\Delta_{\mathrm{max}}^{\mathrm{(SS)}}$ (solid red line); screened in the normal state (NS) $\Delta_{\mathrm{max}}^{\mathrm{(NS)}}$ (dash-dot blue line). 
%Barrier thickness $d =0.7$.
}
 \label{Delta_max}
 \end{figure}
%%%%%%%%%%%%%%%%%%%%%%%%%%%%%%%%%%%%%%%%%%%%%%%%%%

Figure \ref{Delta_max} shows $\Delta_{\mathrm{max}}$ as a function of $r_s$, determined from the three mean-field approaches.  $\Delta_{\mathrm{max}}$ is the maximum value of the momentum-dependent gap $\Delta_{\bm{k}}$ at  zero temperature.  For this example, we give the energy scale of $\Delta_{\mathrm{max}}$ in Kelvin,  taking $m^\star$ and $\kappa$ from double bilayer graphene with a hBN substrate and barrier $d=0.7$.\cite{Perali2013}  While not directly applicable to graphene, since here there is only one valley, we expect the trends to be the same.  

In Fig.\ \ref{Delta_max}, we see that as $r_s$ approaches $r_s=5$, the gap $\Delta_{\mathrm{max}}^{\mathrm{(SS)}}$ calculated with screening in the superfluid state (SS), becomes equal to the gap $\Delta_{\mathrm{max}}^{\mathrm{(US)}}$ without screening (US).  This indicates that (SS) screening is unimportant for $r_s\agt 5$. The absence of screening for $r_s\agt 5$ is caused by the collapse of the Fermi surface in the BEC regime.  Without a Fermi surface there can be no electron-hole excitations that are needed for screening.  The Fermi surface collapse is associated with the renormalization of the chemical potential $\mu$, with $\mu$ going large and negative.  This strong renormalization of $\mu$
is frequently used to characterize the BCS-BEC crossover in ultracold fermionic atoms,\cite{Perali2002} but is less well-known in solids.   
In contrast, the  gap from the normal state screening (NS) approach, $\Delta_{\mathrm{max}}^{\mathrm{(NS)}}$, is very much smaller than $\Delta_{\mathrm{max}}^{\mathrm{(SS)}}$, and the large renormalization of $\mu$ for the (NS) approximation  only occurs at unrealistically low densities, $r_s>15$.

When $r_s$ drops below $r_s=5$, the pairs become less compact, screening starts to become significant, and so  in Fig.\ \ref{Delta_max} $\Delta_{\mathrm{max}}^{\mathrm{(SS)}}$ becomes less than $\Delta_{\mathrm{max}}^{\mathrm{(US)}}$.   By $r_s\alt 3$ we are approaching the BCS regime, where the screening becomes so strong that $\Delta_{\mathrm{max}}^{\mathrm{(SS)}}$ is exponentially suppressed and  drops sharply, before abruptly disappearing  at $r_s=2.5$, leaving only a second exponentially vanishingly small solution.   The overall physical behavior is 
that superfluidity kills screening at low densities, while screening kills superfluidity at high densities. 

The transition to the superfluid state with a large gap is not continuous either in temperature or density.    As a function of  temperature,  the normal to superfluid transition has Kosterlitz-Thouless character.   As a function of density,  the superfluid state at large densities has an exponentially small gap, with sub-mK critical temperatures.   We have noted that at an onset value of $r_s$, a large discontinuous jump in the superfluid gap occurs in the (SS) approach.  This is caused by the sudden appearance of three solutions to the gap equation (Eq.\ \ref{Delta-eqn}) instead of just the one solution.\cite{Lozovik2012}   Only the solution with the largest $\Delta_{\mathrm{max}}$, corresponding to the lowest ground-state energy, will actually be realized in the system, with the result that $\Delta_{\mathrm{max}}$ suddenly becomes large and comparable to the chemical potential $\mu$.   The jump in $\Delta_{\mathrm{max}}$ has a strong discontinuous character, similar to a first order transition.

The quantitative comparisons we have made allow us to also address the role of vertex corrections, which are an important issue in superfluidity for two reasons.  First, we recall that for a Coulomb pairing interaction there is no characteristic energy scale to use in a Midgal expansion,\cite{Migdal} so the beyond-Migdal vertex corrections are not {\it a priori} small \cite{Pietronero} for any of the mean-field approaches, with or without screening. Second, there are additional vertex corrections when screening is evaluated in the superfluid state (SS), arising from the presence of the self-energy insertions in the polarization diagrams needed to generate the  broken-symmetry state.\cite{Ward}
The good agreement  between DQMC and the (SS) approach in Fig.\  \ref{Critical-density}  allows us to conclude that the sum total of the vertex corrections are negligible for the (SS) approach  for $r_s\alt 3$ for all $d$ shown (see also Ref.\ \onlinecite{Lozovik2012}).  
For $r_s\agt 5$, the agreement in  Fig.\ \ref{Delta_max}  between $\Delta_{\mathrm{max}}^{\mathrm{(SS)}}$ and   $\Delta_{\mathrm{max}}^{\mathrm{(US)}}$ indicates that the additional vertex corrections are negligible in the (SS) approach when $r_s\agt 5$, while the reasonable agreement in Fig.\ \ref{Cond-Frac} between the DQMC and (SS) condensate fractions for $r_s\agt 5$, plus the shared flat dependence on $r_s$, indicates that the beyond-Migdal vertex corrections are also small in the (SS) approach when $r_s\agt 5$	.
We thus conclude that  the vertex corrections are small in the (SS) approximation for $r_s\alt 3$ and $r_s\agt 5$, that is, for much of the density range.    
The insignificance of the vertex corrections probably stems from the relatively large number of carrier species in the system and  the opening of a large gap that suppresses particle-hole processes.  

By a similar argument, the agreement between DQMC and the (SS) approach  in Fig.\  \ref{Critical-density}  for $r_s\alt 3$  indicates that the intralayer correlations between like species has little effect on the superfluid properties for $r_s\alt 3$.   This is consistent with conclusions drawn by comparing the gaps reported in Fig.\ 2 of Ref.\ \onlinecite{Zhu95}, which included these correlations, with the gaps calculated in Fig.\ 1 (a) of Ref.\ \onlinecite{Pieri}, which neglected these correlations.  This comparison shows, at most, a $10$-$20$\% effect on the $T=0$ gap.   We note that if bench-mark DQMC data were available for a particular system, then intralayer and
interlayer correlations could be included in a systematic way by adapting the classical-map technique to the superfluid coupling.\cite{CDW}

Having established the (SS) approach as the best approximation for screening in the superfluid state, we now discuss why superfluidity has been so difficult to observe in electron-hole bilayer systems.  
Experiments with  electron-hole monolayers of graphene separated by a barrier thickness of $1$ nm saw no evidence of superfluidity,\cite{Gorbachev} and this is consistent with the theoretical predictions of Ref.\ \onlinecite{Lozovik2012} using the (SS) mean-field approach for  these system parameters.  Reference   \onlinecite{Perali2013} demonstrated, however, that a double bilayer graphene system with a $1$ nm barrier, could generate an exciton superfluid at experimentally attainable temperatures.   Experiments with double quantum wells in GaAs with peak separation of electron and hole wave functions $\agt 25$ nm did not see definitive evidence of superfluidity.\cite{Croxall,Lilly}   This observation is consistent with theoretical predictions  within the (SS) mean-field approach.\cite{Perali2013,private}  Reference \onlinecite{private} showed that carrier densities need to be reduced by a factor of two at existing peak separations, to generate superfluidity at temperatures $T_c\sim 100$ mK.   
Recently, experimental evidence suggesting  existence of preformed electron-hole Cooper pairs in a hybrid graphene-GaAs double layer system with quadratic bands has been reported.\cite{Pelligrini2014}  Below a characteristic temperature, the Coulomb drag  displays an upturn with an order of magnitude enhancement.  The characteristic temperature aligns with the pseudogap crossover temperature, which should be of the order of the pairing energy scale. A fit of the temperature dependence of the drag resistivity gives an estimate of a superfluid  transition temperature 
of $T_c\sim 10$-$100$ mK.  This temperature range is of the same order as the $T_c$ evaluated within the (SS) mean field approach for this system, although at lower densities.\cite{Perali2014} 

In conclusion, we have  resolved a long-standing debate about the best mean-field approach to take for screening in electron-hole bilayer excitonic superfluidity.  We compared DQMC condensate fraction properties with predictions from different mean-field approximations for screening, and we were able to conclude that the best mean-field approximation to use is the (SS) superfluid state screened interaction approach.  The extent of the satisfactory comparison between the DQMC and (SS) results for the condensate fraction over such a wide parameter range cannot be regarded as  fortuitous.  The good agreement for such a fundamental ground state property of the superfluid as the condensate fraction, gives strong support to the predictive power of the (SS) approach, a straightforward theoretical approach based on mean field.  

This agreement of ground state properties should help theoretically in the experimental search of electron-hole superfluidity at accessible temperatures, since it 
now makes it possible to employ the (SS) approximation to explore beyond the practical capabilities of DQMC: to map out finite temperature properties like the superfluid transition temperature $T_c$,\cite{Perali2013} and to investigate new semiconductor and graphene devices with complicated lattice configurations and a large number of Fermion species, all in the quest for high $T_c$.

 \noindent{\bf
Acknowledgements.} We thank Andrew Croxall,  Antonio Castro Neto, Stefania De Palo, Jim Eisenstein, Richard Needs,  Pierbiagio Pieri, Sebastiano Pilati, Marco Polini, Gaetano Senatore, and Inti Sodemann for helpful discussions.

\end{document}